\documentstyle[epsfig]{aipproc}

\begin{document}
\title{The Fluence Duration Bias}

\author{Jon Hakkila$^*$, Charles A. Meegan$^{\ddagger}$, Geoffrey N. Pendleton$^{\dagger}$,\\ 
Robert S. Mallozzi$^{\dagger}$, David J. Haglin$^*$, \& Richard J. 
Roiger$^*$} 
\address{$^*$Minnesota State University, Mankato, Minnesota 56001\\
$^{\ddagger}$NASA/MSFC, Huntsville, Alabama 35812\\
$^{\dagger}$University of Alabama, Huntsville, Alabama 35812}

\maketitle

\begin{abstract}
The fluence duration bias causes fluences and durations of faint gamma-ray bursts to be systematically underestimated relative to their peak fluxes. 
Using Monte Carlo analysis, we demonstrate how this effect explains characteristics of structure of the fluence vs. 1024 ms peak flux diagram.
Evidence of this bias exists in the BATSE fluence duration database, and 
provides a partial explanation for the existence of burst class properties.
\end{abstract}

\section*{Introduction}

The {\it fluence duration bias} is an instrumental bias causing some 
gamma-ray burst fluences and durations to be underestimated relative to 
their peak fluxes. The fluence duration bias does not manifest itself by 
altering the trigger rate, but instead alters measured burst properties. Elsewhere in this conference \cite{hakkila99} we present evidence that 
the class of Intermediate bursts identified by statistical clustering 
analysis \cite{mukherjee98} can be produced from the hardness vs. 
intensity correlation and the fluence duration bias. We also demonstrate 
how the bias can be responsible for decreasing fluences and durations of 
the longest low peak flux Class 1 bursts.
In this paper, we describe the fluence duration bias in more detail.

\section*{An Example}

\begin{figure}[ht!] 
\centerline{\epsfig{file=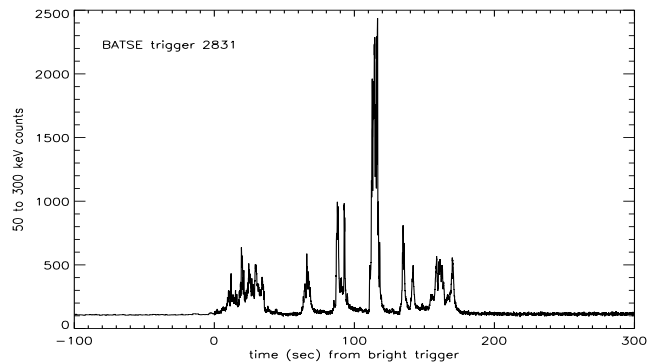,height=2in,width=3.5in}}
\vspace{10pt}
\caption{BATSE trigger 2831.}
\label{fig1}
\end{figure}

Figure \ref{fig1} demonstrates the time history of a bright, Class 1 
(Long) BATSE burst (trigger 2831) as measured in the 50 to 300 keV range 
on the 1024 ms timescale. This burst is complex with 
an overall duration in excess of 180 seconds.

\begin{figure}[ht!] 
\centerline{\epsfig{file=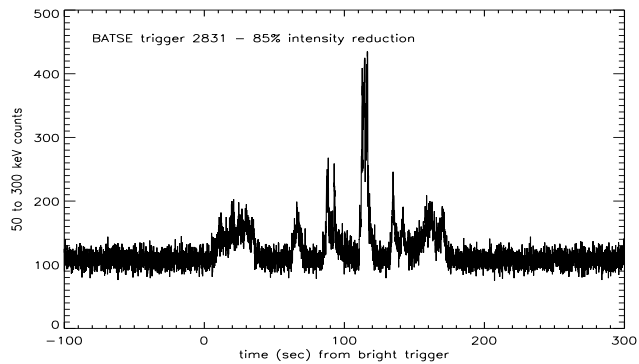,height=2in,width=3.5in}}
\vspace{10pt}
\caption{BATSE trigger 2831, reduced in intensity by 85\%.}
\label{fig2}
\end{figure}

Figure \ref{fig2} is a Monte Carlo simulation of what this burst might 
look like if its 1024 ms peak flux were reduced in intensity to 15\% of 
its measured value (Poisson fluctuations have been added to the reduced 
signal). If the reduced burst duration is assumed to be identical to 
that of the unreduced burst, then its measured fluence-to-peak flux 
ratio is unchanged from the actual value of 19.4 (we measure the result 
in terms of the fluence-to-peak flux ratio, because Poisson fluctuations
can also cause a burst's peak flux to change). If, however, the reduced 
burst duration is determined from ``recognizable pulses'' (pulses that 
are clearly visible above background; our algorithm assumes that the 
first and last peaks larger than $4\sigma$ above background bound the 
burst duration because there is no formal algorithm used by a human 
operator), then the average fluence-to-peak flux ratio drops slightly 
to 94\% of its actual value. 

\begin{figure}[ht!] 
\centerline{\epsfig{file=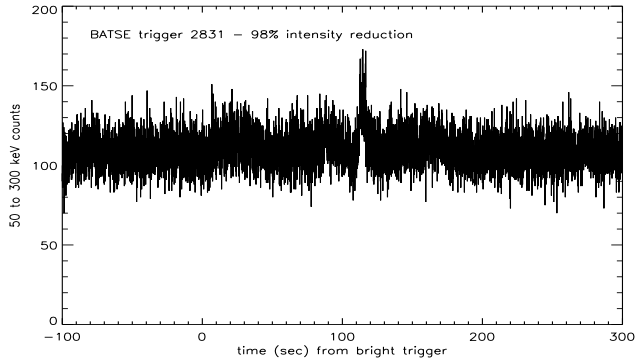,height=2in,width=3.5in}}
\vspace{10pt}
\caption{BATSE trigger 2831, reduced in intensity by 98\%.}
\label{fig3}
\end{figure}

Figure \ref{fig3} shows what the burst might look like if reduced to 
2\% of its actual value. Most of the burst fluence is confined to a 
temporal span of roughly 20 seconds. Our ``recognizable pulse'' 
algorithm finds that the burst is still considerably longer than this 
single pulse, but that the total burst duration is still underestimated 
for the purpose of measuring fluence. The fluence-to-peak flux ratio 
for the burst in question is only 61\% of its actual value. 
This underestimate is even larger when 
the burst is reduced to a value closer to the trigger threshold.

It is difficult to accurately model the process by which the fluence 
duration interval is chosen, since human interaction plays an important 
role. We suspect that the actual amount of the bias is less than the 
amount described here, since the human eye and mind are good at 
removing patterns from noise. Nonetheless, there is evidence that the 
bias is present, and that it is large enough to cause a depletion in 
the number of small peak flux, high fluence bursts as well as being 
responsible for producing some Class 3 burst characteristics from 
Class 1 bursts. 

\section*{Evidence for the Fluence Duration Bias in the 4B Catalog}

Fluence appears to be one of BATSE's most accurately measured 
quantities because its statistical measurement errors are typically 
only $\pm 5\%$. However, there is no intensity-dependent component 
to this measurement error, as might be expected from Figures 
\ref{fig1}, \ref{fig2}, and \ref{fig3}. 
It should be mentioned that there are no BATSE bursts with fluences 
less than zero (as might be expected if background dominated the fluence 
measurement), and few with fluences less than the fluence found in the 
1024 ms peak flux.

\begin{figure}[ht!] 
\centerline{\epsfig{file=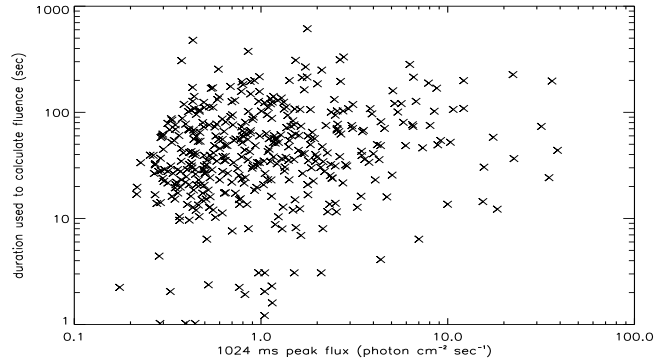,height=2in,width=3.5in}}
\vspace{10pt}
\caption{Fluence Duration vs. 1024 ms Peak Flux from BATSE 4B Data.}
\label{fig4}
\end{figure}

The formal fluence error is kept small in part by fitting the 
background for faint bursts with high-order polynomials. Unfortunately, 
this process can introduce systematic underestimates of burst fluence 
by overestimating background \cite{bonnell97}. {\it The fluence error 
can also be reduced by decreasing the fluence duration}. 
Figure \ref{fig4} plots fluence durations for 
available bursts in the 4B Catalog. The sample has been limited to 
Class 1 bursts detected using the same trigger criteria (because Class 
2 and Class 3 bursts are clearly shorter than the Class 1 bursts, and 
because different trigger criteria might alter the composition of the 
sample in a heterogeneous way).
 
{\it Figure \ref{fig4} indicates that there are few long Class 1 fluence durations near BATSE's detection threshold} (1024 ms peak fluxes slightly greater than BATSE's 0\% efficiency of 0.2 photons cm$^{-2}$ 
second$^{-1}$). This is strong evidence for the existence of the 
fluence duration bias, and it indicates that the magnitude of the 
effect apparently strengthens for fainter bursts.

\begin{figure}[ht!] 
\centerline{\epsfig{file=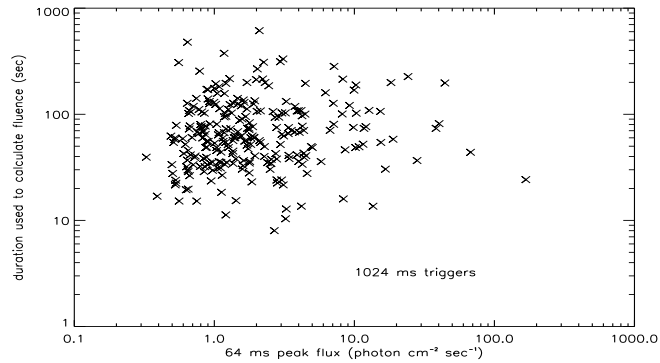,height=2in,width=3.5in}}
\vspace{10pt}
\caption{ Fluence duration vs. 64 ms peak flux for Class 1 bursts triggering on the 1024 ms timescale.}
\label{fig5}
\end{figure}

\begin{figure}[ht!] 
\centerline{\epsfig{file=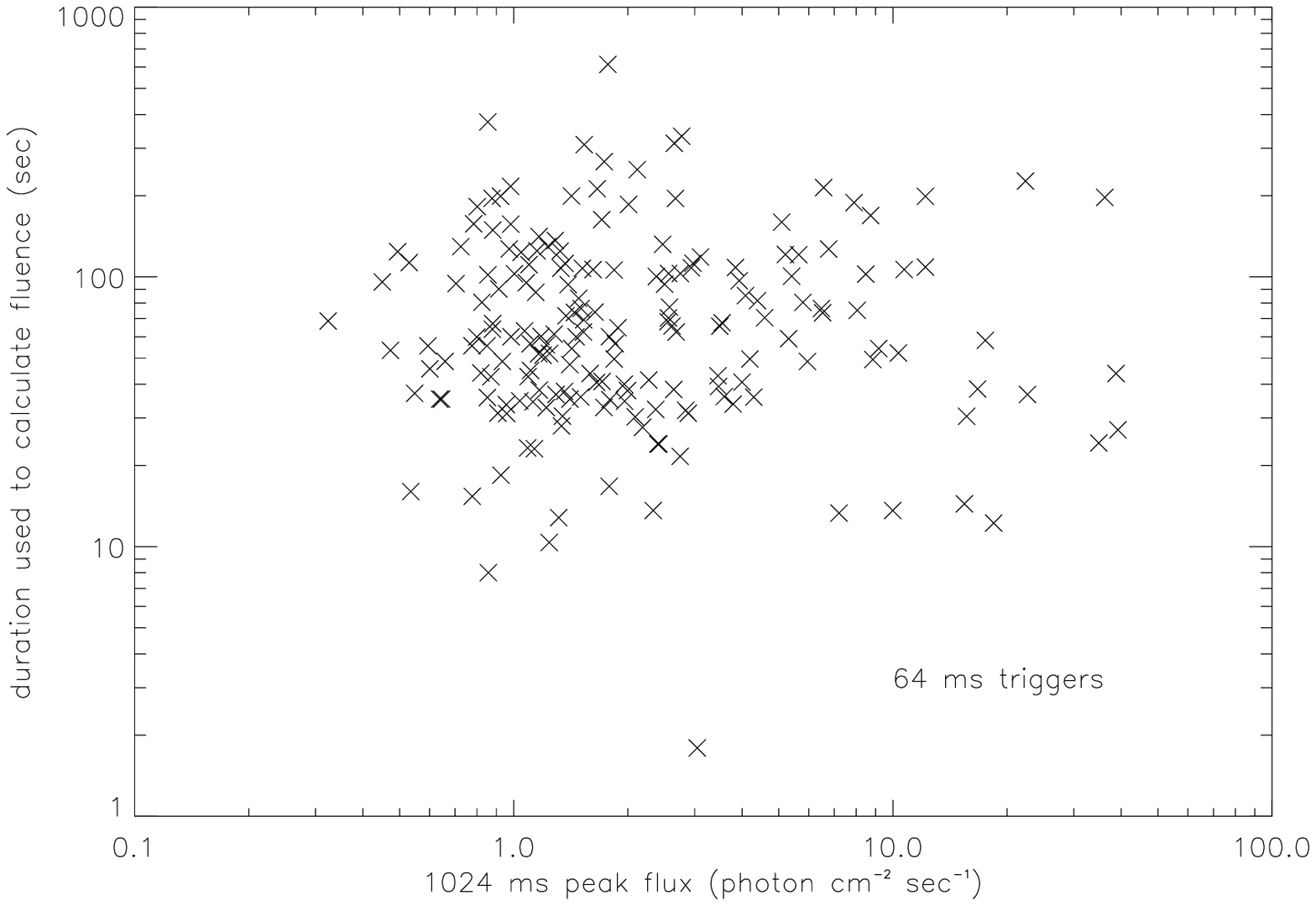,height=2in,width=3.5in}}
\vspace{10pt}
\caption{Fluence duration vs. 1024 ms peak flux for Class 1 bursts triggering on the 64 ms timescale.}
\label{fig6}
\end{figure}

\begin{figure}[ht!] 
\centerline{\epsfig{file=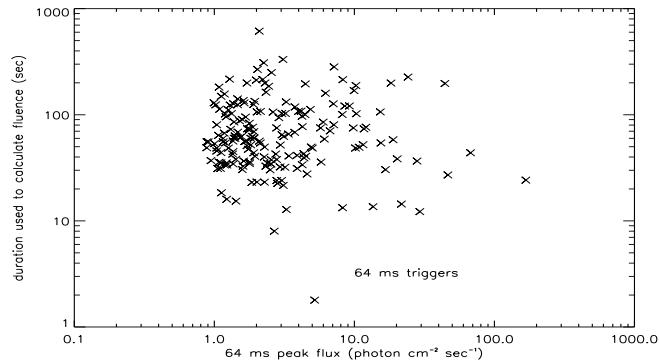,height=2in,width=3.5in}}
\vspace{10pt}
\caption{Fluence duration vs. 64 ms peak flux for Class 1 bursts triggering on the 64 ms timescale.}
\label{fig7}
\end{figure}

Figures \ref{fig5}, \ref{fig6} and \ref{fig7} demonstrate that the 
fluence duration bias is more difficult to cleanly delineate when peak 
flux and/or trigger timescales are shorter than 1024 ms (the effect is 
likewise more pronounced when longer timescales are used). We attribute 
this to the lower signal-to-noise ratio of shorter timescale measurements, 
making intensities measured on these timescales less accurate measures 
with larger intrinsic scatter than 1024 ms. The fluence duration bias 
is still present on shorter timescales; the scatter of these measures
just makes it harder to recognize. 

\section*{Conclusions} 

Monte Carlo modeling of bursts with different temporal structures 
indicates that fluence duration is easy to underestimate, particularly 
for faint bursts. This causes some burst fluences and durations to be underestimated. Some bursts, such as trigger 2831, have temporal 
structures more susceptible to this bias than others. The strength of 
the bias is hard to judge for an individual burst, as it depends both 
on burst temporal morphology and on how the human operator selects a 
fluence duration interval. The magnitude of the bias depends both on 
the time intervals chosen for the peak flux and trigger flux, since the 
fluence underestimate must be made relative to a ``fixed'' brightness 
measure. The fluence duration bias appears capable of producing observed
characteristics of the fluence vs. 1024 ms peak flux diagram, and of making 
some Class 1 bursts (primarily faint ones) take on Class 3 characteristics. 
We currently studying this effect in greater detail.

\end{document}